# Effects of Urban Boundary Layer Turbulence on Firebrand Transport

Iago Dal-Ri dos Santos, Neda Yaghoobian[*]

Florida State University, FAMU-FSU College of Engineering, Department of Mechanical Engineering, Tallahassee, FL, USA

## Abstract

This study investigates the role of topography-induced turbulence, generated by an idealized urban region, in the transport of firebrands and risk of spotting. Flight dispersion, deposition, and smoldering state of tens of thousands of individual mass and size-changing firebrands were investigated in the atmospheric boundary layer turbulence, which was obtained using Large-eddy simulations. Firebrands were assumed to be smoldering spherical particles of Stokes numbers ranging from 30 to 175.

Results indicate that the presence of urban topography significantly affects the firebrand flight behavior, landing distribution, and risk of spotting. Compared to a case with flat topography, horizontal dispersions of the smallest size firebrands were significantly enhanced when urban topography was presented, while the largest firebrands landed closer to each other and closer to the release point. Consequently, a notably different and more compact spotting risk map was achieved. Within the urban boundary layer turbulence, firebrands had a shorter flight and smoldering times in comparison with the flat case. As a result, firebrands landed with larger temperatures, which contributed to a higher risk of spotting in the presence of urban topography.

**Keywords:** Lagrangian Particles, Large-eddy Simulation, Smoldering Particles, Spotting, Urban area, WUI

# 1. Introduction

Spotting is an erratic fire propagation mechanism associated with the transport of burning or smoldering debris by the wind and their subsequent landing. The debris is created from local fuel in the vicinity of the main fire and is lofted to the atmosphere by updraft currents. Once airborne, this debris is carried away by the wind and, on landing, can ignite local fuel far downwind of the main fire front, generating secondary fires known as spot fires. Due to its intermittent nature, the spotting process makes the overall behavior of wildfires and their propagation even more difficult to predict. The irregularity and intermittency of spotting are associated with the condition of turbulence in the boundary layer wind that is modulated by the presence of vegetation, uneven terrain, and/or buoyancy effects. Spotting is also known to be the major mechanism of fire propagation to communities in the wildland-urban interface (WUI) (e.g., Maranghides and Mell, 2011; Manzello and Foote, 2014; Thomas et al., 2017; Urban et al., 2019), and its erratic nature is known to pose serious threats to human lives and properties (Manzello et al., 2008; Maranghides and Mell, 2013; Martin and Hillen, 2016).

The transport of firebrands is characterized by the forces exerted by the local wind on smoldering particles of varying shapes and aerodynamics. As a result, the study of firebrand transport, spotting, and fire propagation requires an understanding of the combinatory factors of the firebrands' temporally variable aerodynamics and thermodynamics together with the spatiotemporally variable structures of atmospheric turbulence. Considering these important factors, the current study aims to focus on an understudied aspect of firebrand transport concerning the role of surface topography and the modulated overlying atmospheric turbulence in the spotting process.

Although still not deeply understood, the transport phase of the spotting process has been studied in several previous works, both numerically and experimentally. In the experimental framework, Tarifa et al. (1967) analyzed the influence of size and shape of spherical, cylindrical, and plate-shaped firebrands on particle trajectories through wind tunnel experiments. Their investigations also included an analysis of mass and size regression rates of the particles. In the works by Manzello et al. (2008, 2009) and Zhou et al. (2015), laboratory experiments were conducted in order to obtain the size and mass distribution of firebrands released under different environmental conditions such as fuel material, moisture content, and wind speed. Analysis of flammability and thermal decomposition of firebrands with different shapes and materials was conducted under laboratory conditions in the work of Ganteaume et al. (2011). Oliveira et al. (2014) performed experimental measurements on the trajectory of non-burning cylindrical particles and proposed a numerical model for calculating the trajectory of firebrands under rotational effects. Song et al. (2017) investigated the effect of wind speed on the transport distance of disk-shaped firebrands and proposed a theoretical heat transfer analysis to explain the deposition behavior. Further, through their wind tunnel experiments, Song et al. (2017) analyzed the transport distance and mass distribution of firebrands and correlated the effect of the wind speed to the particles' landing and extinction behaviors. Tohidi and Kaye (2017a) analyzed both lofting and downwind transport of non-burning rod-like particles in a wind tunnel experiment and statistically described their scattering behavior as a function of flow parameters. Laboratory experiments were also conducted by Suzuki and Manzello (2017), where the accumulation of smoldering firebrands in the vicinity of a flat plate exposed to different wind speeds was studied. They observed that wind speed influences both the spatial distribution and combustion intensity of deposited firebrands. In a more recent study, Suzuki and Manzello (2021) analyzed, both numerically and experimentally, the effect of the separation distance between two rectangular obstacles on the accumulation pattern of smoldering firebrands. Their results indicated that separation distance significantly influenced the accumulation of firebrands.

In addition to laboratory-scale experiments, firebrand characterization has also been performed in prescribed (Filkov et al., 2016) and uncontrolled fires (Rissel and Ridenour, 2013; Suzuki and Manzello, 2018), providing data regarding firebrand size and mass distributions. However, due to the large dimensions and complexities involving full-scale fires and wildfires, it is very difficult to address, in detail, the transport mechanisms through a purely experimental framework (Martin and Hiller, 2016). Therefore, numerical works have been developed alongside experiments in order to supply additional insights into the complexities involved in the firebrand transport phenomena. In the numerical framework, part of the existing models has relied on simplified wind profiles for simulating the transport of firebrands. For example, Fedele (1976) has obtained the probability of spot fire occurrence based on prescribed parameters such as wind speed, fire spread rate, and fuel characteristics. Albini (1979) analyzed firebrand lofting and transport using an analytical plume model under a constant uniform wind velocity. Tse and Fernandez-Pello (1998) developed a numerical model to predict the trajectory of spherical particles carried by a steady logarithmic wind profile. Analytical wind profiles were also employed in the works of Anthenien et al. (2006)

and Albini et al. (2012), where the trajectory of burning particles under the effect of simplified wind-blown fire plumes was analyzed. Martin and Hillen (2016) derived a model for calculating spotting distribution and have modeled wind speed as a function of height using a power-law relation. The main shortcoming of these simplified wind models is the lack of consideration of turbulence effects on the transport of firebrands.

Computational fluid dynamic (CFD) simulations have been employed to account for the effect of turbulence on particle trajectories in firebrand transport. For example, Reynolds Averaged Navier-Stokes (RANS)-based models have been employed to analyze the spreading and deposition of firebrands under different topographical and wind conditions (e.g., Huang et al., 2007; Koo et al., 2012). However, RANS methods are appropriate for analyzing the effect of the mean flow, as they do not provide information on the turbulent velocity fluctuations. Large-eddy simulations (LES), which can accurately resolve the main turbulence structures, have, therefore, been used. For example, Himoto and Tanaka (2005) employed LES to analyze the trajectory of disk-shaped particles across a turbulent boundary layer flow under the buoyancy effects. LES is also employed in the work of Thurston et al. (2017) where the landing distribution of firebrands of negligible inertia and constant fall velocities was analyzed with respect to different buoyancy-driven flows. Tohidi and Kaye (2017b) used LES wind velocity data in a stochastic particle transport model to investigate the landing probability of firebrand particles. LES has also been employed by Anand et al. (2018) to analyze the scattering behavior of non-burning cylindrical particles, while spherical burning particles were analyzed in the LES works of Bhutia et al. (2010) and Pereira et al. (2015).

As mentioned earlier, this study aims to investigate the transport of firebrands and the spotting process under the effect of topography-modulated turbulence. In general, regarding the effect of surface topography on firebrand transport, to our best knowledge, only a few studies can be identified: Hilton et al. (2019) computationally investigated the effect of two hill-shaped topographies on the path of spherical firebrands and their depositions and have shown that the presence of obstacles introduces additional turbulent structures, which significantly modify the firebrand dispersion behavior. In addition, the previously mentioned experiments from Suzuki and Manzello (2017, 2021) have analyzed the impact of a flat plate and two rectangular obstacles on the propagation of smoldering firebrands under different wind speeds, finding that the modified wind structures play an important role in the propagation behavior of firebrands. None of these studies considers the effects of relevant large-scale turbulence structures that form within the atmospheric boundary layer. In the urban microclimatology community, it has been shown that such large-scale structures are significantly important in the transport phenomena near the earth surface and within surface-mounted obstacles (e.g., Kanda et al., 2004; Coceal et al., 2007; Inagaki and Kanda, 2010; Takimoto et al., 2011; Inagaki et al., 2012; Perret and Savory, 2013).

Despite that the spotting process is an important fire propagation mechanism in WUIs, there is little information available regarding firebrand transport in urban areas. Due to the presence of building structures, urban areas create microclimates and local flow fields that are significantly different than other areas on the earth's surface. Review of the literature indicates that currently there is no fundamental studies of the effect of topography-induced flow structures on the firebrand behavior and spotting risk that could be extended to urban and WUI areas. This work aims to computationally investigate the effect of topography-induced flow structures on the trajectory, landing distribution, and spotting risk of smoldering firebrand particles in an idealized urban region. For this purpose, LES was employed and firebrand particles were assumed to be spherical-shape smoldering (temperature-, mass-, and size-changing) particles. Observational evidence shows that in nature, firebrands can be of any random shape and the shape of firebrand particles influences the aerodynamic forces, and thus the flight behavior of the particles and their subsequent landing distributions (e.g., Sardoy et al., 2007; Sardoy et al., 2008; Koo et al., 2012; Oliveira et al., 2014). Selection of canonical-shaped spherical (e.g., Lee and Hellman, 1970; Tse and Fernandez-Pello, 1998; Bhutia et al., 2010), cylindrical (e.g., Albini et al, 2012; Oliveira et al, 2014; Anand et al., 2018), and disk (e.g., Himoto and Tanaka, 2005; Anthenien et al., 2006; Sardoy et al., 2008) particles for firebrands makes the complexity of the spotting problem simplified for computational studies and allows for introducing identical base situations throughout the literature when fundamental problems are investigated.

The rest of the paper is structured as follows: Sect. 2 describes the models used in this study and the simulation setup. In Sect. 3, the results relating to the flow and particle behaviors are presented. While the focus of this study is on the turbulent transport of spherical particles, the landing distribution of cylindrical and disk-shaped particles in an idealized urban setting was also briefly discussed and compared with that of the spheres at the end of Sect. 3. Conclusions and final remarks are presented in Sect. 4.

## 2. Model Description and Simulation Set-Up

In this work, LES is employed to obtain an accurate state of the atmospheric boundary layer turbulent flow over a terrain representing an idealized urban region composed of uniformly distributed cubical structures. For comparison purposes, similar simulations have been performed for a case with flat terrain. The collected instantaneous turbulent flow field then is used to feed a Lagrangian particle tracking model for spherical smoldering firebrands. When relevant, firebrand dynamics in a no-wind flat-terrain case is also investigated. The LES and the firebrand models are described below.

**2.1. Model Description**

*2.1.1 LES-based modeling of the atmospheric wind flow*: For computational modeling of the atmospheric turbulence, the PALM Model (Raasch and Schröter, 2001; Maronga et al., 2015; Maronga et al., 2020) was employed. Through this model, the LES modeling of the turbulent flow is carried out by solving the non-hydrostatic, filtered Navier-Stokes equations under the Boussinesq approximation. The turbulence closure is based on a 1.5-order closure model after Deardorff (1980) for parametrizing the subgrid-scale covariance terms. Time-stepping is performed using a third-order Runge-Kutta scheme, while advection terms are discretized using a fifth-order upwind scheme after Wicker and Skamarock (2002). Near the wall, momentum fluxes are parameterized using the Monin-Obukhov similarity theory. Additional details related to the PALM model and its formulation can be found in Maronga et al. (2015) and Maronga et al. (2020). PALM has widely been used for computational investigations of atmospheric boundary layer and urban flows and different aspects of it have been extensively validated (e.g., Letzel et al., 2008; Park et al., 2012; Letzel et al., 2012; Park et al., 2013; Yaghoobian et al. 2014; Lo and Ngan, 2015; Duan and Ngan, 2019; Duan et al., 2019).

*2.1.2 Aero-thermodynamics modeling of spherical firebrands*: For modeling the firebrand dynamics, particles are tracked individually across the flow, and their three-dimensional (3D) positions and velocities are obtained by solving the equation of conservation of linear momentum in a Lagrangian frame of reference according to: $m_p \, dV_p/dt = F_d + F_g$. In this formula, $m_P$ and $V_P \, (= dr_p/dt)$ are the firebrand mass and velocity vector, respectively, with $r_p$ being the position vector of the particle. $F_g$ and $F_d$ are the gravitational and drag forces, respectively, defined as:

$$F_d = \frac{1}{2} C_d \, \rho_{air} \, A_p |V_r|^2 \frac{V_r}{|V_r|} \tag{1}$$

$$F_g = (\rho_p - \rho_{air}) \forall_p \, g \tag{2}$$

where $V_r \, (= V_w - V_p)$ represents the 3D relative velocity vector between the firebrand velocity ($V_p$) and the velocity of the wind local to the particle ($V_w$) that is obtained from the LES. $C_d$ is the drag coefficient, $A_p \, (= \pi d_p^2/4)$ is the particle's projected area with $d_p$ being the particle diameter, $g$ is the gravitational acceleration, $\forall_p$ is the volume of the particle, and $\rho_p$ and $\rho_{air}$ are the densities of the firebrand and air at 25°C, respectively. Since in this work particles are assumed to be perfectly spherical and non-rotational, there will be no Magnus effect (Mehta, 1985) acting on the firebrands and therefore, no lift force is considered in the particle trajectory model. The drag coefficient ($C_d$) in Eq. 1 is based on an empirical relation for a smooth sphere (Clift and Gauvin, 1970), as adopted for the spherical firebrand trajectory modeling in Anthenien et al. (2006):

$$C_d = \frac{24}{Re_d}(1 + 0.15 Re_d^{0.687}) + \frac{0.42}{(1 + 4.25 \times 10^4 Re_d^{-1.16})} \qquad Re_d < 3 \times 10^5 \tag{3}.$$

In Eq. 3, $Re_d$ is the Reynolds number based on the firebrand diameter. The released firebrands are large enough so that additional force components (i.e., Basset force and Saffman lift) are assumed to be negligible in comparison with drag and gravitational forces (Wang and Squires, 1996). Particle time-stepping is conducted using the second-order Runge-Kutta Predictor-Corrector scheme with the timesteps satisfying the Courant–Friedrichs–Lewy (CFL) condition based on the particle velocity and the LES minimum grid size. This condition ensures that the particle timesteps are small enough to allow the particle dynamics at each timestep to be calculated based on the data within the spatial resolution of the LES.

In order to account for changes in firebrands' mass and size due to pyrolysis effects, following Tse and Fernandez-Pello (1998) and Anthenien et al. (2006), a uniform radial regression model is considered. The model is steered by a single parameter that controls both the size and mass regression rates of each firebrand. This parameter ($\beta$), called the regression coefficient, is based on an experimentally correlated equation presented by Williams (1985):

$$\beta = \beta_o(1 + 0.276Re_d^{0.5}Pr^{1/3}) \tag{4}$$

in which $\beta_o$ is the burning coefficient for oak wood under no wind and $Pr$ is the Prandtl number of the surrounding air. Following Tse and Fernandez-Pello (1998), the density of char is assumed to be small and therefore is negligible compared to that of the unburned material. The diameter of the pyrolysis front, $d_{pyr}$, and the external diameter of the particle that includes the charred portion, $d_P$, are then calculated as

$$\frac{d(d_{pyr}^2)}{dt} = -\beta \tag{5}$$

$$\frac{d(d_P^4)}{dt} = -\chi\beta^2 t \tag{6}$$

where $t$ is the flight time and $\chi = 2\sqrt{3}$ is a fitting coefficient. Firebrands are assumed to be made of rigid oak wood for which the thermodynamic properties are obtained from Tse and Fernandez-Pello (1998). Following Tse and Fernandez-Pello (1998), firebrands burn until their mass reaches 24% of their initial mass at the release point (Tarifa et al., 1967), after which the pyrolysis process ceases, and the particles lose their heat to the environment. To model the temperature evolution of each individual firebrand over time, the transient energy budget equation is solved for each particle based on the lumped capacitance assumption (Tse and Fernandez-Pello, 1998):

$$\frac{dT_p}{dt} = -\frac{S_p}{(\rho \forall c)_p}\left[\bar{h}(T_p - T_\infty) + \sigma\varepsilon(T_p^4 - T_\infty^4)\right] \tag{7}.$$

In this equation, $T_p$, $S_p$ ($= 4\pi r^2$), $\rho$, $\forall$, $c$, and $\varepsilon$ are the temperature, surface area, density, volume, specific heat, and emissivity of the firebrands, respectively. $\bar{h}$ ($= k_{air}\overline{Nu}d_p^{-1}$) is the average convective heat transfer coefficient based on the average Nusselt number $\overline{Nu}$ ($= 2 + 0.6Re^{1/2}Pr^{1/3}$), obtained from the correlation given by Ranz and Marshall (1952).

A firebrand is assumed to be landed once its centroid reaches the location of a surface (obstacle or ground). Upon landing, firebrands are assumed to be immediately attached to the surface, and therefore do not roll or slide across it. It is also assumed that the lifetime of the firebrands ends on landing. Thermodynamic properties of air are acquired at a film temperature, which is defined as the arithmetic mean of the particle surface temperature and the ambient temperature. Variations of air properties with respect to the instantaneous film temperature are accounted for using equations obtained from polynomial regression taken from the literature (Incropera and DeWitt, 1990). The instantaneous velocity of the flow local to the particle ($V_w$) is attained from the LES computational grids at the instantaneous location of the flying particle. This is done by linear interpolations of the 3D velocities in time and space using the time separation between the flow snapshots and the spatial resolution of the LES data. The input flow field consists of snapshots of the instantaneous velocity components ($u$, $v$, and $w$), separated by an interval of 0.5 s. The cumulative volume of the flying firebrand particles to the volume of the fluid domain is small ($\sim 10^{-10}$), and therefore a one-way coupling between the LES and the particle models is considered appropriate (Elghobashi, 1994; Kuerten, 2016).

*Validation of the firebrand trajectory and burning model*: A validation study was conducted to evaluate the accuracy of the firebrand model and its formulation for both charring and non-charring particles. This validation study was performed using the model setup and results from Tse and Fernandez-Pello (1998), on which a major part of the model is based. The main model setup and assumptions are described below, followed by the validation results of the current model. Trajectories of rigid spherical firebrands of different diameters were obtained under the effect of a steady logarithmic wind profile. Particles were released from a fixed point, 10 m above a flat, horizontal surface. The logarithmic wind profile was based on the following equations, with the wind velocity being fixed to 48.3 km h$^{-1}$ at the release height.

$$V_w(z) = \frac{V_*}{\kappa} \ln\left(\frac{z}{z_0}\right), \text{ for } z \geq z_0;$$

$$V_w(z) = 0, \text{ for } z < z_0,$$

(8).

In Eq. 8, $V_*$ is the friction velocity, $\kappa$ (= 0.4) is the von Kármán's constant, and $z_0$ is the roughness height (being equal to 0.05 m). The firebrand trajectories were obtained for wooden (oak wood, $\rho_p = 545$ kg m$^{-3}$) and metallic (aluminum, $\rho_p = 2702$ kg m$^{-3}$) particles. Particles were released while burning with a fixed temperature of 993 K for oak wood and 2753 K for aluminum. Additional information on the thermophysical properties of oak wood and aluminum can be found in Tse and Fernandez-Pello (1998). Wooden particles produce char until extinction, which happens after their mass reaches 24% of their initial values. Upon extinction, wooden firebrands lose heat to the environment, and their temperature decreases according to Eq. 7. On the other hand, metallic particles do not produce char and are allowed to burn until their mass is completely consumed.

Wooden particles are released with a zero initial velocity, while metallic particles are ejected with a horizontal velocity of 1 m s$^{-1}$. Figures 1a and 1b compare the firebrand trajectories for the charring wooden and non-charring aluminum particles against those found in Tse and Fernandez-Pello (1998), while Fig. 1c compares the mass regression results for the wooden particles.

To validate the burning model, Tse and Fernandez-Pello (1998) used the experimental results of Tarifa et al. (1967) for wooden spheres burning in forced convection. In their work, they compared the instantaneous values of the mass and surface area of the firebrands with those obtained experimentally. These instantaneous values of mass and surface area were normalized by their respective values prior to burning (i.e., $m_0$ for the mass and $A_0$ for the surface area). In their comparison, a sphere with an initial diameter of 0.022 m was exposed to steady wind speeds of 8 m s$^{-1}$ and 12 m s$^{-1}$. For brevity, only the comparison for the 12 m s$^{-1}$ wind is shown here. The burning coefficient was assumed to be $\beta_0 = 5.3 \times 10^{-7}$ m$^2$ s$^{-1}$ and a fixed surface combustion temperature of 993 K was considered for oak wood. Figure 1d compares the regression of the normalized mass ($m/m_0$) and area ($A/A_0$) of the burning spheres with those found in Tse and Fernandez-Pello (1998) and Tarifa et al. (1967). The results indicate the accurate performance of the model.

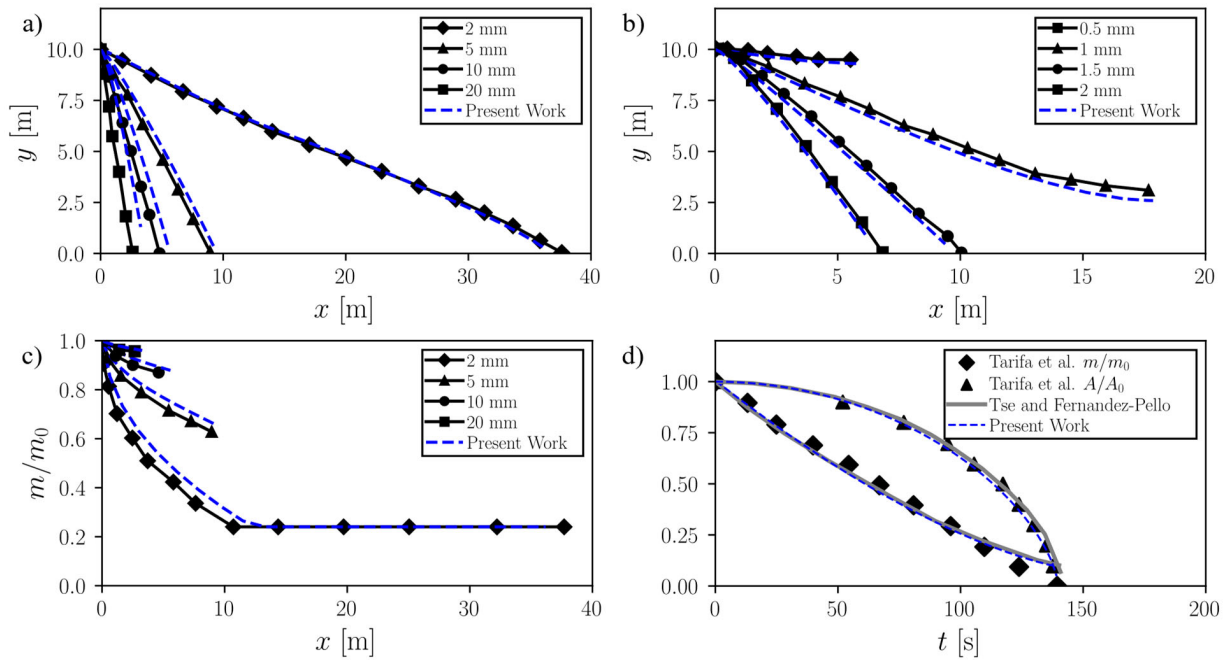

Figure 1: Comparison of the particle trajectory for (a) wooden (charring) and (b) metallic (non-charring) spheres, and (c) mass regression of wooden spheres against the results from Tse and Fernandez-Pello

(1998). (d) Mass and surface area regression for a wooden sphere ($d$ = 0.022 m) under a 12 m s$^{-1}$ wind compared against the results from Tse and Fernandez-Pello (1998) and Tarifa et al. (1967).

## 2.2 Simulation Set-Up

Two simulation domains were considered: 1) a domain consisting of an array of surface-mounted cubical obstacles representing an idealized 3D urban region, and 2) a domain consisting of a flat surface. In the non-flat case, the cubes have square footprints of length and height $H$ = 19.84 m and are equally spaced by the same distance $H$ in both streamwise and spanwise directions. It should be noted that such topography is commonly used in the urban microclimatology community for representing idealized urban regions in fundamental studies (e.g., Kanda et al., 2004; Kim and Baik, 2010; Inagaki et al., 2012; Yaghoobian et al., 2014; Guo et al., 2020; Chen et al., 2020; Chen et al., 2021, Saxena and Yaghoobian, 2022). A schematic of the computational domain of the non-flat case is shown in Fig. 2. In the figure, a representative 'spanwise canyon' and a representative 'streamwise street' that will be referred to on several occasions in the paper are shown in red and blue shaded colors, respectively.

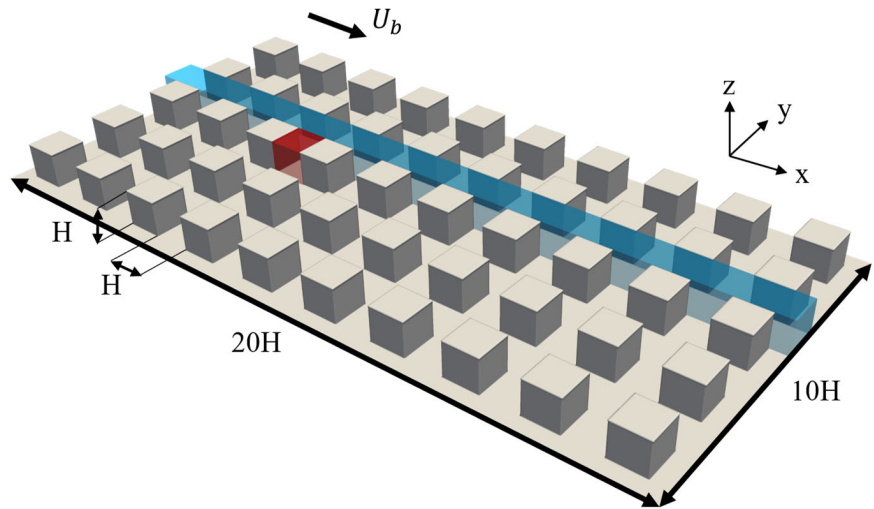

Figure 2: Schematic overview of the simulation domain for the non-flat case. The red and blue shaded areas, respectively, represent one of the spanwise canyons and one of the streamwise streets. The mean wind direction is parallel to the x-axis of the domain shown by the vector $U_b$.

To estimate an appropriate domain size, it was ensured that the domain is horizontally large enough to contain the most energetic eddies passing over the obstacles and vertically heigh enough to capture the effect of large turbulent coherent structures that form aloft within the inertial sub-layer of the boundary layer. Therefore, integral length scales of the turbulent eddies in the domain (which provide a measure of the extent of the region over which the velocities are correlated), as well as different mean flow properties (e.g., streamwise velocities, turbulent kinetic energy (TKE), and velocity variances) across the domain were compared between several domains of different lengths, widths, and heights. The integral length scale increases with the increase in the domain extent up to a point beyond which the increase in the domain size has little or no effect on the integral length scale. Our analyses indicated that a domain size of $20H$ (length) × $10H$ (width) consisting of 10 by 5 uniformly spaced cubes/buildings results in an integral length scale of 38.3 m. This domain size is over ten times larger than its associated integral length scale, which is beyond the minimum eight-times criteria required for isotropic turbulence (Pope, 2001). The height of the domain ($20H$) was chosen large enough to ensure that the effects of the turbulent coherent structures within the inertial sublayer on the near-wall flow is considered (e.g., Finnigan, 2000; Watanabe, 2004; Coceal et al., 2007; Inagaki et al., 2012). Such a design corresponds to an urban topography with a frontal area density ($\lambda_f$) and a plan area density ($\lambda_p$) both equal to 0.25. For determining an appropriate grid resolution for the simulations, different flow parameters were compared between simulations with grid sizes of $0.015H$ and $0.03H$. The comparison results indicated that the flow features are significantly close to each other. Therefore, the grid size of $0.03H$, corresponding to 0.62 m, was selected for the simulations, which gives a resolution of 32 grid points across each cube side. Grid stretching in the vertical direction is employed beyond height $3H$, with an expansion factor of 1.08 and a maximum element height of 5.5 times the minimum grid size.

Cyclic boundary conditions are applied at the lateral limits of the domain, with a zero gradient Neumann condition applied at the top boundary and a no-slip condition at the bottom and over the obstacles' surfaces. The flow was stirred with a mean streamwise wind velocity of 18.8 m s$^{-1}$. Using this setup, the simulations were run for over 300 eddy turnover times (based on the domain size and mean velocity) until the flow reached a fully developed steady turbulent sate as indicated by the convergences of the flow statistics. After this stage, the firebrand particles were released in the domain. The bulk wind speed of 18.8 m s$^{-1}$ is based on the average measured wind speeds reported from wildfires in New Jersey (obtained from Anand et al. (2018)) which is consistent with measurements reported from other large wildland fires (Cruz et al., 2020). No fire spread behavior was modeled in the present work, and therefore fire-atmosphere interactions and the related buoyancy effects are not considered. All simulations were conducted under isothermal conditions (i.e., neutral atmospheric stratification) to ensure that the flow statistics purely represent mechanically induced flow fields. The simulation setup for the flat topography uses similar characteristics to that of the non-flat case, only missing the surface-mounted obstacles.

## 2.3 Suites of Simulations

In order to analyze the effect of topography over the firebrand transport, statistical distribution of landing positions, together with the evolution of first and second-order statistics along firebrands' trajectories are obtained for both flat and non-flat cases. Firebrands were released from a single fixed point within the flow field, located at $x$ = 5 m, $y$ = 60 m (midpoint of the y-axis), and $z$ = 50 m (~ 2.5$H$). As observed in previous works (e.g., Tohidi and Kaye, 2017a; Anand et al., 2018), our analysis (briefly discussed in Sect. 3.5) also indicated that the particle deposition statistics are significantly sensitive to the particle release height ($\delta$), which is directly related to the permanence time of the particle in the turbulent flow (i.e., the higher the release height, the longer the flight time through the flow). However, in this work, the particles' release height (as well as the initial particle velocity) is fixed in order to isolate the effect of the topography-induced flow on firebrand dispersion and to reduce the number of free parameters in the modeling procedure. This assumption allows for a consistent comparison between the two topography cases, as it guarantees that the differences in particle dispersion observed are driven exclusively by differences in the flow structures. Particles are all released with a fixed initial streamwise velocity equal to the average freestream wind velocity in the domain (i.e., 18.8 m s$^{-1}$). A large enough number of particles (minimum of 50,000) were released for each case in this study in order to achieve statistical convergence of mean and standard deviation of the particle landing positions, selected through a sensitivity analysis.

In this work, the diameter of the released spherical firebrands varies randomly (in a uniform distribution) from a minimum of $d_p$ = 2 mm up to a maximum of $d_p$ = 5 mm. Since LES is employed, the particles receive information of the resolved flow. Under certain conditions, however, the unresolved subgrid-scales (SGS) structures can significantly impact particle behavior (Kuerten, 2016; Marchioli, 2017). In order to estimate the effect of such structures over the particle trajectory, the particle relaxation time $\tau_p$ is calculated and compared against the characteristic flow timescale $\tau_\Delta$. The ratio between these two timescales yields the Stokes number $Stk = \tau_p/\tau_\Delta = \rho_p d_p^2/[18\mu(\Delta^2/\epsilon)^{1/3}]$, in which $\rho_p$ and $d_p$ are the particle's density and diameter, respectively, $\mu$ is the dynamic viscosity of air, $\Delta$ is the LES characteristic grid size, and $\epsilon$ (= $U_b^3/L_c$) is the energy dissipation, estimated based on the integral length scale of the flow in the LES domain ($L_c$). For the parameters employed in the present study, $Stk$ ranges between 30 ($d_p$ = 2 mm) and 175 ($d_p$ = 5 mm), which falls within the limit in which particles are assumed to be SGS-inertial (i.e., unaffected by the SGS turbulence) and are influenced mainly by the resolved scales of the flow (Marchioli, 2017). Our analysis indicated that the mass- and size-changing particles remain in the SGS-inertial range throughout their flight.

## 3 Results and Discussion

### 3.1 Flow characteristics

Figure 3 shows snapshots of the instantaneous turbulent flow velocity magnitudes (normalized by the bulk velocity, $U_b$) for both flat and non-flat topography cases. The snapshots are taken from a vertical cross-section at the middle of the spanwise axis (y = 5$H$), providing visual representations of the different flow fields in the two cases. It is evident that in the urban topography case, the velocity magnitude is significantly smaller closer to the obstacles, and recirculation regions form within the spanwise canyons (more visible in mean flow figures (not shown)). The wake produced by the obstacles propagates within the domain, creating a skimming flow regime (Oke, 1988) over the obstacles while enhancing turbulence within the flow aloft. A quantitative comparison between the flow characteristics of the two topography cases is presented in Fig. 4, where temporally (over 1800 s) and horizontally averaged vertical profiles of normalized streamwise velocity and TKE are shown. From Fig. 4a, a decrease in the mean streamwise velocity is evident within and above the urban canyon layer (UCL), while the enhancement in the mean TKE is observed in the same region of the non-flat case (Fig. 4b). The mean TKE has a peak at the obstacle height, which is due to the presence of the cavity shear layer forming at this location (e.g., Letzel et al., 2008). The interactions between the flow and the obstacles create additional turbulent structures, which propagate throughout the domain and contribute to the increase in the boundary layer height. Such differences in the flow structure between the two topography cases drive significantly different particle behavior and spreading, which will be discussed in detail in the following sections.

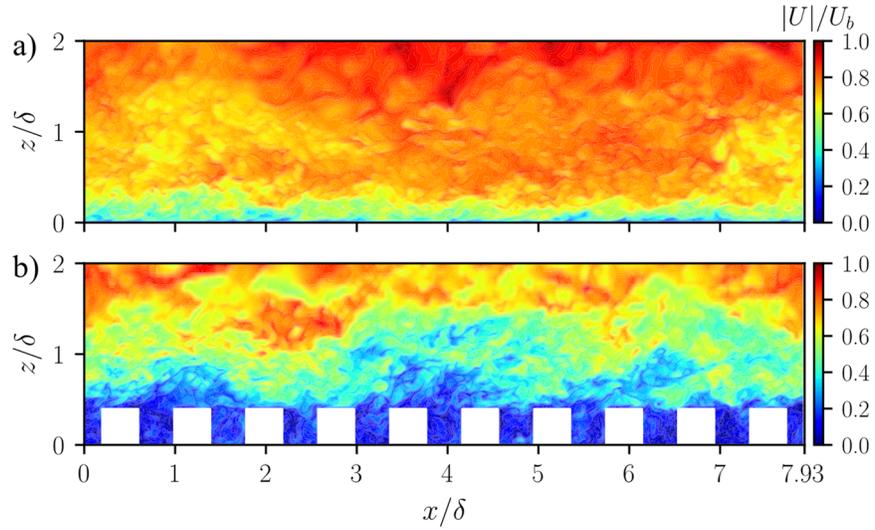

Figure 3: Snapshots of the instantaneous velocity magnitudes for the (a) flat and (b) non-flat topography cases from a vertical plane at the center of the domain's spanwise axis. The axes are normalized by the particle release height ($\delta$), and the snapshots are shown only up to twice $\delta$.

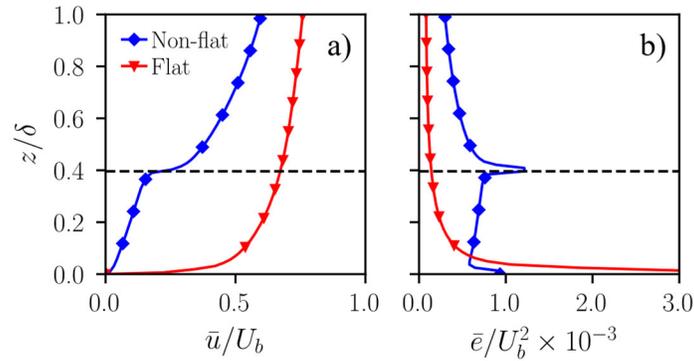

Figure 4: Comparison of temporally and horizontally averaged vertical profiles of normalized (a) streamwise velocity and (b) TKE for the flat and non-flat cases. Profiles are shown up to $z/\delta = 1$. The height of the obstacles is demarked by a dashed line.

### 3.2 Firebrand deposition characteristics

The firebrand deposition patterns and their propagation distances are analyzed and compared between the flat and non-flat cases. Previous works in the literature regarding the behavior of inertial particles in turbulent flows (e.g., Wang et al., 2006; Mollicone et al., 2019) have shown that particle statistics is a strong function of particle inertia. To represent the effect of particle inertia, the following results and discussions are presented in terms of discrete ranges of firebrand diameter with corresponding Stokes numbers as follows: 2-3 mm ($Stk \sim 30 – 60$), 3-4 mm ($Stk \sim$ 60 – 110), and 4-5 mm ($Stk \sim 110 - 175$). Figure 5 compares landing positions of firebrands of different sizes using different colors on the ground surface between the two topography cases. The figure only shows the region of the domain where the majority of firebrands are landed.

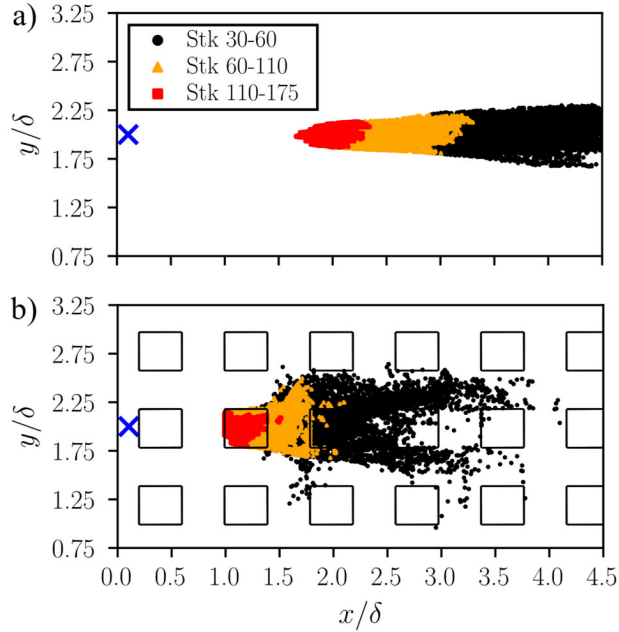

Figure 5: Landing positions of firebrands of different sizes for the (a) flat and (b) non-flat cases. Spatial coordinates are normalized by the firebrand release height. The release position is marked by a blue cross.

From Fig. 5, it can be seen that the particle landing distributions are notably different between the two topography cases. The presence of obstacles influences the deposition behavior of firebrands in three distinct ways: first, a significant portion of firebrands collide with the obstacles before reaching the ground level, causing their landing positions to be closer to the release point. Second, firebrands propagating across the UCL are exposed to weaker streamwise wind velocities (Fig. 4a), which contributes to the reduction in particle propagation distance in the non-flat case. Lastly, firebrands experience stronger turbulence (i.e., larger flow fluctuations) in the non-flat case (Fig. 4b), which causes the particles to disperse more strongly in the spanwise direction. It should also be noted that as the smallest firebrands enter the under-roof level flow (i.e., the UCL), they tend to disperse strongly within the streamwise streets rather than flying over the buildings, creating elongated streaks along the streets. Such behavior is attributed to the larger flow velocity within the confined areas of the streamwise streets compared to the flow above the buildings.

The larger firebrands, having larger inertia, are more resistant to the influence of the turbulence while the smallest particles are more susceptible to the turbulence motions and tend to disperse further over the horizontal plane, indicating that the dispersion mechanism is a strong function of firebrand size.

Analyzing in more detail the deposition patterns of firebrands, Fig. 6 presents the probability density function (PDF) and spanwise standard deviations of firebrand landing positions along the streamwise direction. These results present the statistical behavior of deposited firebrands shown in Fig. 5 in quantitative terms. The PDFs are obtained using a Gaussian kernel with a bandwidth calculated using Scott's rule (Scott, 1992), while the spanwise standard deviation is calculated discretely based on the spanwise mean position of firebrands landing within discrete intervals of $0.1\delta$ in the streamwise direction.

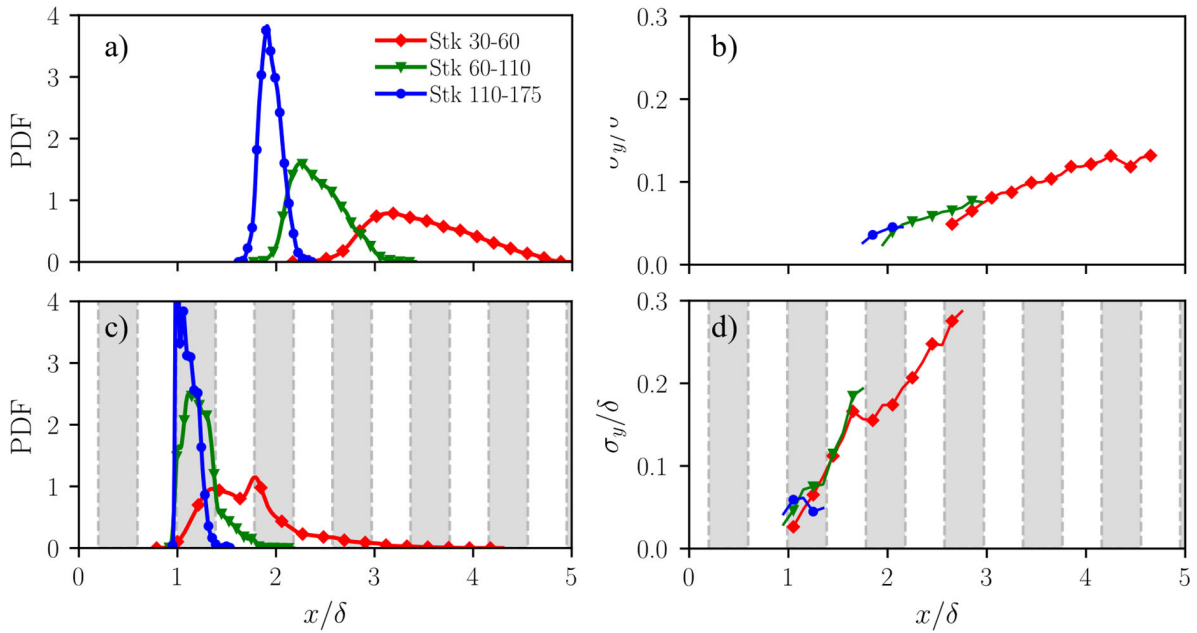

Figure 6: Probability density function and spanwise standard deviation of landing positions along the streamwise direction for the flat (a and b) and non-flat (c and d) cases. The locations of the obstacle columns are shown with gray color bars.

In accordance with the results in Fig. 5, the PDF distributions in Fig. 6a and 6c show an increase both in streamwise landing dispersion and landing distance from the release point with the decrease in firebrand size. Besides landing closer to the release point, the larger size firebrands land closer to each other, as can be seen from the reduced spreading about the mean landing position. These behaviors are more pronounced in the non-flat case. For the flat topography, statistics of the landing positions, especially for the small and medium-size firebrands, resemble a Rayleigh distribution along the streamwise direction, showing a rapid increase in concentration closer to the release point, followed by a gradual reduction downstream. Such spreading distribution over flat topographies is also reported in previous numerical (Wang, 2011) and experimental (Zhou et al., 2015) works regarding firebrand deposition.

In the non-flat case, both the largest and intermediate size firebrands concentrate mainly over an obstacle in the second column (Fig. 6c and 5). There are two spikes in the PDF of the largest firebrands. The initial spike, located at the same position as the frontal face of the obstacle, indicates that a significant portion of the largest firebrands are deposited in this location, with the remaining of them landing over the rooftop of the same obstacle. Within the spanwise canyon located about $x/\delta = 1.5$, the intermediate size firebrands show reduced concentration with the increase in distance (which is not clear from Fig. 5). The PDF of the smallest firebrands shows a distinct bimodal distribution around the $x/\delta = 1.5$ canyon, with its largest peak occurring at the windward face of an obstacle in the third column of buildings. The less pronounced peak of this PDF happens at the leeward face of an upstream obstacle. From Figs. 6b and 6d, it can be seen that the spanwise dispersion of the firebrands increases with the propagation distance, which is more pronounced for the smallest and intermediate size particles in the non-flat case.

### 3.3 Impact on smoldering and flight time

It is of interest to also understand the impact of turbulence and topography on the ignition potential of firebrands. Previous studies in the literature (e.g., Manzello et al., 2008; Ganteaume et al., 2011) have described firebrand ignition potential in terms of the average individual firebrand mass, size, and temperature. In the present study, a dimensionless thermal energy parameter ($E^*$) is defined that considers the thermal energy of each deposited firebrand and the firebrand concentration over the surface. This allows us to analyze the distribution of spotting potential across the bottom surface of the domain. For the calculation of $E^*$ (Eq. 9), the ratio between the thermal energy of each individual firebrand on landing ($m$) and on release ($m_o$) is calculated ($m^* = m/m_o$). Then, the firebrand concentration on the ground is calculated as the number of firebrands landing over patches of arbitrary size. The average thermal energy per patch is then calculated to reflect the effect of particle accumulation.

$$E^* = \frac{1}{N_i} \sum_{A_i} c^* \times m^* \times T_p^*, \qquad (9)$$

In Eq. 9, $N_i$ is the number of deposited firebrands over each patch $i$ of area $A_i$ and $c^* = c/c_0 = 1$ is the firebrand specific heat capacity on landing ($c$) with respect to that at the release point ($c_o$), assumed to remain unchanged. $T_p^*$ is the dimensionless temperature that is defined as $T_p^* = (T_p - T_\infty)/(T_o - T_\infty)$, with $T_p$, $T_0$, and $T_\infty$ being the firebrand temperature on landing, the firebrand temperature on release, and the ambient temperature, respectively. Equation 9 does not account for other known local parameters that influence ignition potentials, such as fuel moisture, fuel composition, and air moisture, and it is meant to indicate the ignition potential purely due to the firebrands' thermal energy and accumulation at landing. In these analyses, patches are defined by uniformly dividing the bottom of the domain into discrete square areas of size $1/8H$. This rather arbitrary size was selected as a good compromise between the resolution and the capability to capture firebrand clustering regions over the domain. All firebrands landing within the horizontal limits of these patches are accounted for.

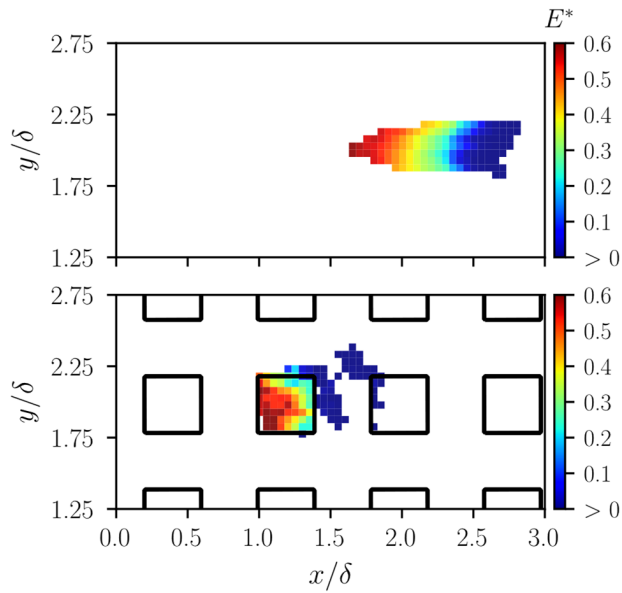

Figure 7: Distribution of dimensionless thermal energy for the (a) flat and (b) non-flat cases.

Figure 7 shows the distribution of the thermal energy of the landed firebrands. From the definition of $E^*$, the contours in Fig. 7 represent the areas with the risk of spotting. It can be seen that in both topography cases, the regions with the highest risk coincide with regions in which the largest firebrands are deposited (Fig. 5 and 6). In the flat case (Fig. 7a), the distribution of spotting risk follows that of the firebrand distribution on the ground surface: An area of large ignition potential is present closer to the release point, where the firebrands with larger mass, and therefore larger energy, are deposited. The ignition potential then decreases gradually along the x-axis, as both the firebrands' deposited mass and concentration on the ground decrease. For the non-flat case (Fig. 7b), however, a non-uniform $E^*$ distribution is observed with the highest potential being associated with the location where the majority of large and midsize firebrands are landed (e.g., over the building in the second column; Fig. 6c). At the ground, the ignition potential is significantly reduced in both magnitude and area covered when compared with the flat case. The ground-level non-zero ignition potential regions are located in areas where higher firebrand concentrations for small and intermediate size firebrands are observed.

The thermal energy of the firebrands on landing depends on the influence of turbulence on the flight time and the smoldering state of firebrands during their flight. Figure 8 compares the ensemble-averaged firebrands' smoldering and flight times between the two topography cases, where the ensemble average is evaluated by averaging these times for the firebrands from an individual size range.

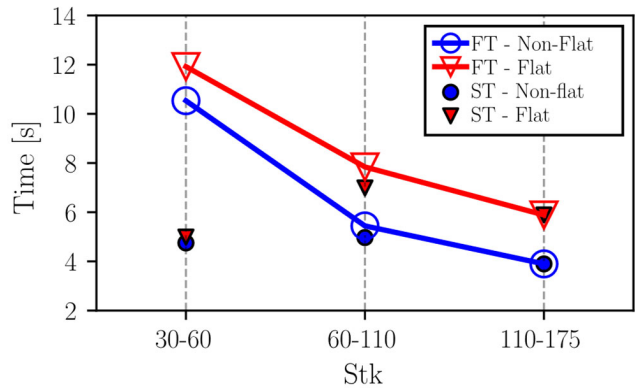

Figure 8: Comparison of ensemble-averaged smoldering (ST) and flight (FT) times for different firebrand size ranges between the flat and non-flat topography cases.

Figure 8 shows that, in general, firebrands have shorter flight times in the non-flat case, which are a result of the lower flow velocity that carry the firebrands and the collision of the firebrands with the obstacles. As expected, larger firebrands have smaller flight times due to their larger weight, but they have higher smoldering times due to their larger mass. For the largest firebrands (Stk 110

– 175), however, the smoldering time is the same as the flight time, as these particles land before reaching the end of their smoldering process. In addition to this, these large firebrands are less susceptible to the effect of turbulence structures and tend to conserve a relatively stable trajectory throughout their flight time. The largest firebrands are deposited across a smaller area and, thus, contribute to the largest risk of spotting (as discussed before).

Smoldering times in the non-flat case are also smaller in comparison with the flat case, more noticeably for the intermediate and large size firebrands. This behavior is, again, driven by the firebrands' collision with the obstacles prior to reaching the ground level and the assumption that the firebrands' lifetime ends as soon as they land. As a consequence of the reduced flight times, these collisions increase the fraction of the released firebrands that land while smoldering. The smaller difference between smoldering and flight times also indicates that extinguished firebrands are deposited with large temperatures, which contributes to the higher risk of spotting in the non-flat case.

### 3.4 Firebrand trajectory characteristics

To understand the temporal evolution of Lagrangian statistics of flying firebrands, quantitative information on firebrands' behavior during their flight is analyzed. This information is presented as instantaneous ensemble-averaged statistics of flying firebrands with respect to the firebrand flight time. The ensemble averaging is done using linearly interpolated position and velocity data of firebrands at discrete intervals of 0.1 s of flight time. Only flying firebrands at each instant are considered in the calculations of the ensemble average, which is performed until the flight time reaches the mean flight time value of each size range. For results in which multiple firebrand sizes are considered, all flying firebrands at each instant are considered in the ensemble average and the mean flight time of the smallest size firebrands is used as the stopping criteria

Figure 9 presents first-order statistics of the streamwise and vertical ensemble-averaged positions and velocity components of all flying firebrands. First-order statistics of the spanwise components are not shown as, despite large variations in the cross-stream direction, the mean values remain close to zero. These statistics are compared between the two topography cases (flat and non-flat) and a case with no wind under a flat terrain. The latter is used as a reference to isolate the effect of inertia-driven dispersion and to help understand the extent to which local turbulence influences firebrand statistics.

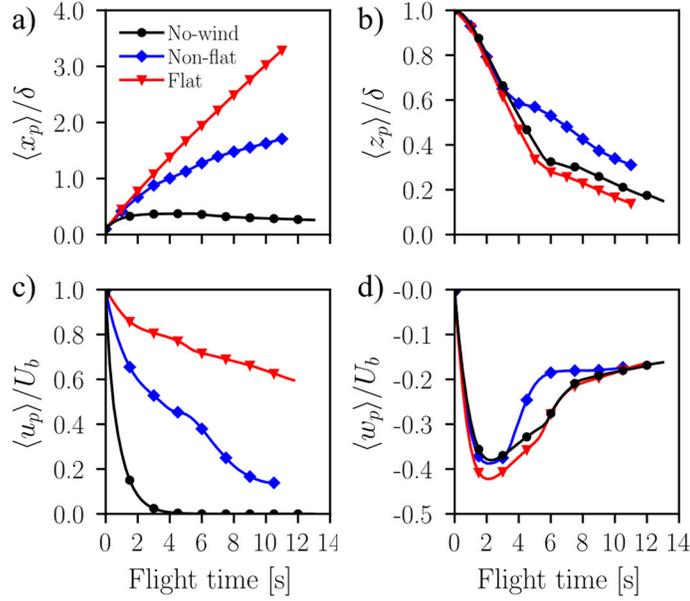

Figure 9: Instantaneous ensemble-averaged a) streamwise and b) vertical positions, and c) streamwise and d) vertical velocity components of flying firebrands of all sizes as functions of firebrand flight time. Positions and velocities are respectively normalized by the firebrand release height and flow bulk velocity.

Figure 9a indicates that firebrands in the non-flat case have smaller streamwise propagation during their flight in comparison with those in the flat case. As discussed earlier, this behavior is partly attributed to the reduced mean streamwise velocity within the boundary layer over the obstacles. Within the weak-velocity urban boundary layer region (Fig. 2), the contributing drag (Eq. 1) is smaller, leading to smaller particle streamwise velocities as seen in Fig. 9c. However, in comparison with the no-wind case, both streamwise propagation (Fig. 9a) and velocity (Fig. 9c) in the flat and non-flat cases are much larger. For the no-wind case, the negative streamwise drag acting on the firebrand as it crosses the stagnant air is much stronger, and the firebrand velocity quickly decays along its trajectory. From such decay in streamwise velocity, the firebrands reach an almost purely vertical trajectory around the second 3 of the flight time, as indicated by the very small streamwise velocities in Fig. 9c.

Figure 9d shows that the firebrands' vertical velocity $\langle w_p \rangle$ converges towards a constant value with the increase in the flight time. This value corresponds to the average terminal velocity of the ensemble firebrands. From Fig. 9d, it can be seen that firebrands in the non-flat case achieve this terminal velocity faster in comparison with the flat case. This behavior and, in general, the behavior of vertical velocities of the flying particles do not appear to have a direct relationship with the flow characteristics, mainly, for three reasons: i) although the terminal velocity is achieved faster in the non-flat case, there is a convergence to a similar value with that obtained in the flat cases; ii) the time in which an increase occurs in the vertical velocity of the non-flat case (i.e., second 3) coincides with the mean flight time of the largest firebrands that collide with the obstacles before landing. This means that statistics of the largest particles are not accounted for in the ensemble averaging after 3 s of the flight time. This is expected to cause a decrease in the average vertical velocity profile since the remaining firebrands are lighter. This also induces the observed increase in the vertical position of particles in the non-flat case at around the collision time (Fig. 9b); and iii) a comparison between the behavior of ensemble-averaged particle vertical velocities between the two flat cases (with and without wind) indicates that for the ranges of particle stokes number studied here, $\langle w_p \rangle$ is barely influenced by turbulence, being only slightly larger in the case with the wind.

In order to better understand the differences in the firebrand behavior during their flight as a result of topography-induced turbulence, the temporal evolution of firebrand turbulent dispersion along the particles' trajectory is analyzed. Such analysis provides a quantitative comparison between the two topography cases and helps identify regions of flow in which firebrand dispersion may be enhanced or restrained. To perform these analyses, first, the instantaneous displacement of each firebrand is calculated as the distance of the firebrand from the ensemble average position of all firebrands flying at an instant $t$, following:

$$x'_p(t) = x_p(t) - \langle x_p(t) \rangle, \tag{10}.$$

In this equation, the angled brackets represent the ensemble average of the firebrands in flight. The instantaneous (turbulent) firebrand dispersion is then calculated as the ensemble-averaged variance of the firebrand instantaneous displacement, $\langle x'_p(t) x'_p(t) \rangle$. Figure 10 shows the results for firebrand turbulent dispersion in horizontal (*x* and *y*) and vertical (*z*) directions as functions of the flight time.

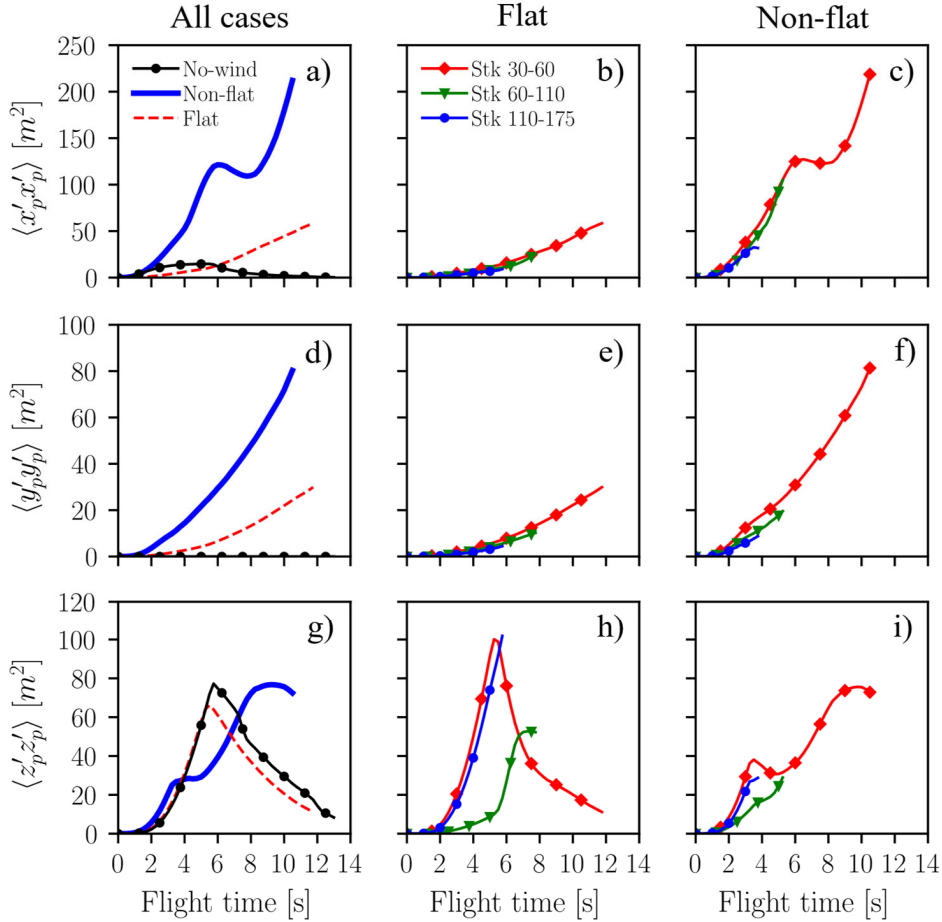

Figure 10: Comparison of the instantaneous ensemble-averaged firebrand dispersion along a) $x$, d) $y$, and g) $z$ directions between all cases, and between different firebrand sizes for the flat (b, e, and h) and non-flat (c, f, and i) cases as functions of firebrand flight time.

A comparison of horizontal turbulent dispersion of ensemble firebrands between the three (flat, non-flat, and no-wind) cases (Figs. 10a and 10d) indicates that firebrands in the non-flat case deviate substantially from their ensemble mean position during the flight compared to the other cases. From Fig. 10a, it can be noticed that in the non-flat case, the streamwise dispersion has a distinct reduction, which starts at around the second 6 of flight time, roughly matching the mean flight time of the intermediate size firebrands, as indicated in Fig. 10c. The period of the decrease spans around the second 6 to second 8 of the flight time and corresponds to the period when a large portion of small firebrands (Stk 30 – 60) collide with the third column of obstacles and settle (Fig. 5). After this reduction, the streamwise turbulent dispersion substantially increases as the remaining flying firebrands are those with smaller inertia that largely fly through the higher wind velocity regions of the streamwise streets. At this stage, the firebrands propagating along the streamwise streets are less susceptible to collisions with obstacles, and the dispersion increases monotonically, similar to the behavior observed for the flat case.

When comparing the turbulent dispersion behavior between the flat and no-wind cases, it is observed that both streamwise and spanwise dispersion components (Fig. 10a and 10d) are significantly enhanced under the presence of turbulent flow. Indeed, for the case without wind, spanwise dispersion is non-existent since firebrands are released with zero spanwise initial velocity. This indicates that trajectory deflections caused by turbulent gusts from surrounding flow are the main reason why firebrands disperse horizontally. The vertical dispersion component (Fig. 10g), however, seems to be only slightly affected by turbulence when comparing the flat and no-wind cases. This suggests that the turbulent flow does not influence vertical dispersion as much for the range of firebrand sizes considered. For large Stokes numbers, vertical dispersion of particles is mainly inertia-driven, meaning that gravitational settling is the main mechanism in which firebrands disperse vertically. It is expected that smaller firebrands to be more affected by turbulence-induced vertical fluctuations: In Fig. 10g, a peak is observed in the vertical dispersion at around the second 5 of flight time, for both flat and no-wind cases, that is followed by a significant reduction. As seen in Fig. 10h, this behavior is driven mainly by the smallest size firebrands. As the firebrands land, the average vertical displacement of the remaining flying firebrands in relation to the ensemble mean decreases, causing the observed reduction in dispersion. Such behavior is not observed for the largest firebrands, and it is just slightly observed for the intermediate-size firebrands. Because of their relatively short trajectories, larger firebrands land with relatively small variations in position and flight time, and the reduction in firebrand displacement as they land is less pronounced. For the non-flat case, the two regions of reduced vertical dispersion (at seconds 3 and 9; Fig. 10g) are, respectively, induced by the deposition of the largest and smallest size firebrands. The behavior of the horizontal turbulent dispersion for individual group size of firebrands in the flat case (Fig. 10b and 10e) is qualitatively similar to those found by Anand et al. (2018) for non-burning cylindrical particles, showing an increase in turbulent dispersion with flight time and the dominance of streamwise dispersion over the spanwise dispersion.

A well-documented behavior of inertial particles in turbulent flows is the ejection of these particles out of vortical regions of flow (e.g., Wang et al., 2006; Bhatnagar et al., 2016). When these particles are trapped in vortical regions, they obtain an outward motion induced by their inertia that causes an increase in particle concentration in the outer region of vortical structures. However, this behavior is known to happen at a specific range of Stokes number for which particles have enough residence time in the circular motions (Bhatnagar et al., 2016). As reported by Wang et al. (2006), this behavior is found to cause a reduction in the average vorticity magnitudes felt by the particles. In order to analyze the presence of such behavior for the firebrands (that have variable mass and size throughout their lifetime) and the general impact of vortical motions of flow on the firebrand trajectory, the instantaneous ensemble-averaged TKE and vorticity magnitudes felt by flying firebrands are investigated (Fig. 11). This instantaneous ensemble-averaged data is evaluated at discrete intervals of 0.1 s of the flight time.

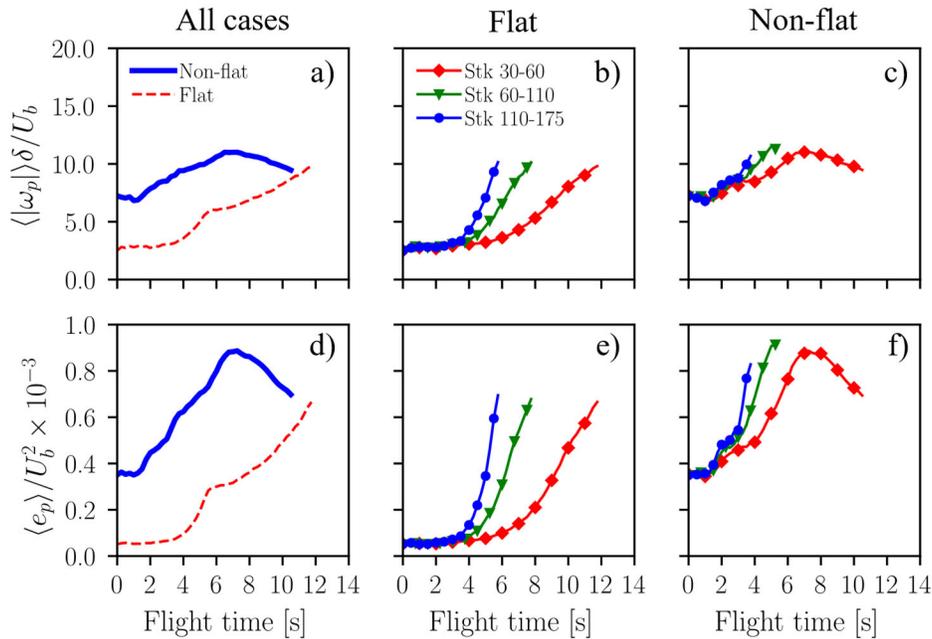

Figure 11: Top row: instantaneous ensemble-averaged normalized vorticity magnitude felt by flying firebrands between all cases (a), and for different firebrand sizes for the b) flat and c) non-flat cases. Bottom row: instantaneous ensemble-averaged normalized TKE felt by flying firebrands between all cases (d), and for different firebrand sizes for the e) flat and f) non-flat cases.

As expected, Fig. 11a and 11d show an increase in both TKE and vorticity magnitudes felt by the ensemble firebrands during their flight in the non-flat case in comparison with the flat case. For the flat case, a monotonic increase in both TKE and vorticity profiles is observed with the increase in the flight time. This behavior, which applies to all firebrand sizes, complies with the expected trend observed in turbulent flows over flat surfaces, in which both TKE (outside of the viscous sublayer) and vorticity increase closer to the wall. Also, results for different firebrand sizes in the flat case (Figs. 10b and 10e) indicate that, for the same flight times, particles of different sizes are exposed to different values of vorticity and TKE. In these results, the curves appear to be shifted towards larger flight times with decreasing firebrand size. This behavior indicates that, in the flat case, it takes longer for the smaller firebrands to cross different regions of the flow in comparison with larger firebrands. However, besides the visible shift, no significant differences in the trend and maximum values of the curves are visible for the flat case. This behavior suggests that the firebrands, throughout their size-changing lifetime, are still larger to have enough residence time in the vortical structures and concentrate in regions of low vorticity, and thus, they do not show a visible reduction in mean vorticity over time.

In the non-flat case, both TKE and vorticity felt by the ensemble particles decrease at around the second 7 of the flight time (Figs. 11a and 11d). This decrease is due to the reduced TKE and vorticity felt mainly by the small size (Stk 30 – 60), and slightly by the intermediate size (Stk 60 – 110) firebrands (as seen in Figs. 11c and 11f). As seen in Fig. 4, the flow in the non-flat case has a region of low TKE below the obstacle height within the UCL. Since the major portion of the largest firebrands land on top of an obstacle and do not reach the low-TKE region below the obstacle height, a monotonic increase can be observed in the profile of TKE felt by these large particles (Fig. 11f). For the smallest firebrands, most of which cross the UCL, the effect of the lower TKE felt by the flying particles is visible. Indeed, the maximum values for both TKE (Fig. 11f) and vorticity (Fig. 11c) felt by the intermediate size particles are similar to those for the smallest firebrands. However, since a significant portion of the intermediate size firebrands lands on top of the obstacles, their shorter flight time, and thus their effect in reducing the ensemble-averaged flight time, is sufficient to make the TKE reduction less noticeable in Fig. 11f.

**3.5 Effect of firebrand shape on the firebrand landing distribution**

Firebrand particles can assume a wide range of shapes and sizes (Filkov et al., 2016; Suzuki and Manzello, 2019) depending on the type of fuel, flame intensity, and other environmental factors. These shapes may deviate significantly from the typical canonical shapes considered in the literature, which can create additional aerodynamic effects and lead to differences in firebrand trajectory, combustion, and deposition behaviors. Modeling of such effects is non-trivial and beyond the scope of this work. However, the effect of particle shape on the firebrand landing distribution was briefly investigated here by comparing the deposition characteristics of disk and cylindrical-shaped firebrands with those of spherical firebrands in the same idealized urban setup. The equations for tracking the non-spherical particles follow the work of Anthenien et al. (2006). The correctness of the numerical model based on these formulations (presented in the Appendix section) was evaluated by comparing the firebrand trajectory, mass loss rate, and size regression against the same in Anthenien et al. (2006).

Like the spherical firebrands, uniform combustion along the radius of the firebrands was assumed for both cylinder and disk-shaped particles. The formulation assumes that the cylinders fall perpendicular to the relative wind, and the effects of the oscillatory motions are neglected. For heavy particles (for which density is much larger than the air density), the oscillatory motions tend to lower the effective drag coefficient (Chritiansen and Barker, 1965), which, on average, leads to a reduction in the particle travel distance. As discussed by Anthenien et al. (2006), these reductions are within 20% of normally reported drag coefficients for steady cylinder motions and, therefore, are neglected. In a similar way, the oscillatory motions of disk-shaped firebrands were not considered. Instead, disks were assumed to fall with their symmetry axis being parallel to the relative wind, which maximizes the aerodynamic drag induced by the surrounding flow. Following Anthenien et al. (2006), cylinders with aspect ratio of 3 and disks of a small aspect ratio of $10^{-1}$ were considered. Such a small aspect ratio for disks leads to a very small disk thickness. Combined with the assumption of pure radial size regression, the charred region produced in disk-shaped firebrands is assumed to be significantly thin, so it detaches easily during the particle flight and, therefore, was neglected in the drag coefficient calculation. For consistency in the comparison study, the production of char is also neglected for spheres and cylinders.

Figure 12 compares the landing distributions of the spherical, cylindrical, and disk-shaped firebrands for different firebrand sizes. The results are presented in terms of firebrand diameters that are ranged between 2 – 5 mm (typical sizes that are considered in the literature, e.g., Anthenien et al. (2006), Albini et al. (2012), Pereira et al. (2015), Hilton et al. (2019)). The same release conditions were applied to all firebrands, with the release height being 40 m in these analyses, lower than what is considered before. The lower release height allowed most of the disk-shaped firebrands, which showed large downstream propagations, to land within the computational domain and far from the right boundary, while making the computational costs affordable. Although this reduction causes a reduction in the overall dispersion distance of all particles, quantitative comparisons with previously presented results for $\delta = 50\ m$ indicate that the main deposition features are preserved for both flat and non-flat cases.

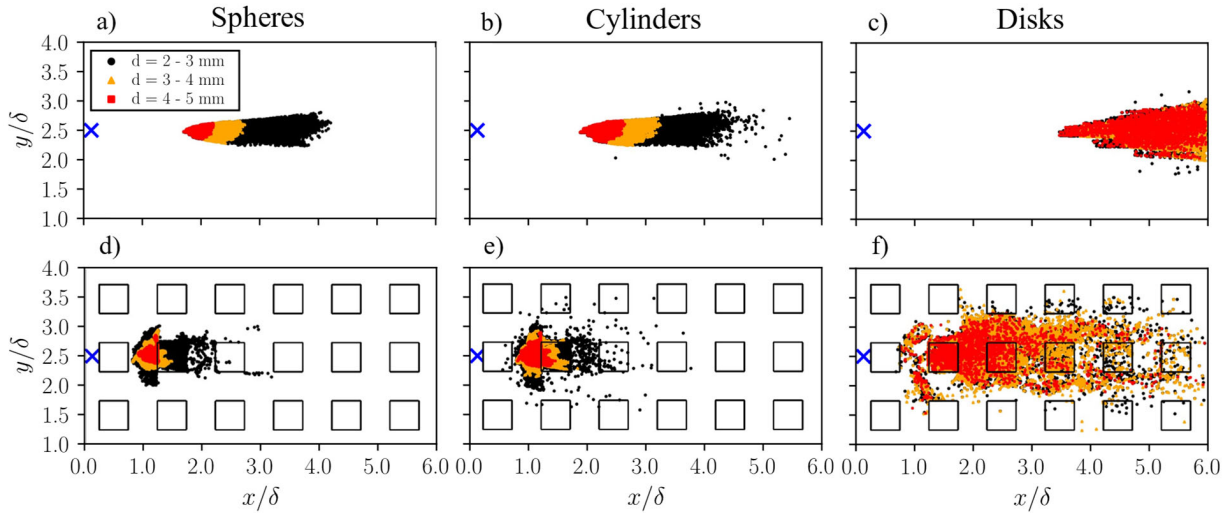

Figure 12: Landing positions of firebrands of different sizes and different shapes for the flat (a, b, c) and non-flat (d, e, f) cases. Spatial coordinates are normalized by the release height $\delta$. The release position is marked by a blue cross.

From the results presented in Fig. 12, similarities between the overall spreading behavior of spheres and cylinders are observed, with cylinders presenting a slightly larger propagation in both horizontal directions. This is due to the smaller mass-to-area ratios of cylinders in comparison with spheres, which leads to larger mean drag-to-weight ratios and makes the cylinders to be more susceptible to the local turbulence fluctuations. In addition to that, as mentioned by Anthenien et al. (2006), cylindrical firebrands have larger surface areas compared to the spherical firebrands, which leads to faster combustion and larger rates of mass loss. As cylinders lose mass faster than the spherical particles, the role of the instantaneous flow force on them throughout their flight is higher than their inertia forces, leading to larger spreading, especially, at the final stages of the combustion and trajectory when compared with the spherical firebrands of the same initial diameters.

Disk-shaped firebrands show the largest average spreading and propagation distances between all shapes. Compared with the spheres and cylinders of the same diameter, these particles present the smallest mass-to-area ratios, leading to the largest propagation distances and spreading. Another feature observed from the deposition results of the disks is a smaller effect of particle size on particle trajectory and landing distribution when compared with cylindrical and spherical firebrands as a more homogeneous distribution of firebrand sizes is observed in this case (Fig. 12c,f). Under the small aspect ratios of the disks, changes in radius lead to a smaller impact on the firebrand mass and, therefore, on their contributing gravitational forces in the particle trajectory. Concerning the effect of topography on the firebrand transport, results from Fig. 12 indicate that the presence of topography influences the overall spreading behavior of different shape firebrands in similar ways. For example, as was seen before, in the presence of obstacles, particles of all shapes land closer to the release location compared to their corresponding flat cases. In addition, when the condition (release height and aerodynamics of particles) is suited for the particles to have enough travel time before landing (e.g., in these simulations, in the case of the disk-shape firebrands), regions with a larger concentration of deposited firebrands are observed along the streamwise canyons; the similar feature that was observed previously for the spherical firebrands in Fig. 5. However, when favorable conditions are not supported (here in the case of spherical and cylindrical particles), their overall landing distribution compared to their corresponding flat cases are similar. The mechanisms that induce such changes in spreading behavior are the same as the ones described in detail in Section 3.2.

## 4. Conclusions

In this study, the effect of topography-induced turbulence on the deposition, dispersion, and spotting risk of smoldering firebrands was analyzed to provide a deeper understanding of the transport of firebrands over urban regions. The turbulent flow was simulated over a flat terrain and a terrain composed of an array of cubic obstacles using LES. A Lagrangian particle model was employed to obtain the trajectory, landing positions, and temporal evolution of smoldering spherical firebrands in response to the effects of turbulence. Firebrands of different sizes were released from a fixed point within the boundary layer, and their statistics were compared and analyzed. In addition to spherical particles, the effect of topography-induced turbulence on the landing distribution of smoldering cylindrical and disk-shaped firebrands was investigated.

The results indicated that the presence of turbulent wind significantly enhances the horizontal components of firebrand turbulent dispersion in comparison with a case without wind. Such enhancement is induced by turbulent gusts acting on the firebrands traveling through the boundary layer flow; an effect that is more pronounced for smaller particles.

Due to the presence of obstacles and their role in weakening the near-surface wind flow, firebrands deposited in the non-flat topography case had smaller propagation distances and smaller streamwise dispersion compared to those in the flat case. However, due to enhanced turbulence, firebrands experience larger spanwise landing dispersion in the non-flat case. This behavior was observed for all firebrand sizes and shapes analyzed, although differences in average propagation distance and spreading were evident between the spherical and non-spherical firebrands. These differences are driven mainly by the variations in the mass-to-area ratio between different firebrand shapes. Firebrands with smaller mass-to-area ratios (e.g., disks) propagate and spread further due to their larger drag-to-weight ratios and larger fluctuations in drag force induced by the local turbulence.

The presence of topography also affected the spotting risk significantly, which is directly related to the distribution/accumulation of deposited firebrands and their thermal energy at the landing. It was observed that regions of large spotting risk are located where the largest firebrands land, as these firebrands carry larger thermal energy and have smaller landing distribution. In general, larger spotting potentials happen in the non-flat case. This increase is driven by the reduced firebrands' average flight time, leading to an increase in the number of firebrands deposited while smoldering and an increase in the average temperature of deposited extinguished firebrands.

For the non-flat case, two separate areas of non-zero spotting potential were observed at the ground, indicating regions of preferential concentration of the deposited firebrands. However, preferential concentration regions obtained in the present work are less pronounced than those previously reported in the literature regarding particle deposition over obstacles. The reason is that the firebrands in the present study have larger inertia and resist the turbulent motions more intensely.

The findings of this study contribute to a better understanding of the role of surface topography in the atmospheric boundary layer on spotting and fire propagation risk, especially in wildland-urban interface areas. For a better understanding of the fundamental physics of firebrand transport over complex terrains, it is desired to extend the study to other idealized urban topography scenarios. It is also desired to extend the studies to non-canonical random shape particles and topographies. In addition, understanding the behavior of firebrands' motion after landing requires extended investigations, especially in scenarios involving complex terrains. Particles, after landing, will likely experience additional motions (e.g., rolling and rebounding) prior to reaching their final resting positions. The accurate prediction of the settling behavior requires accurate modeling of the complex interactions between the particles, surface, and local terrain.

# Acknowledgment

This work is supported by funding from the Florida State University and National Science Foundation (NSF) Grant Number CBET 2043103.

# Appendix: Numerical models for cylinder and disk-shaped firebrands

For tracking the cylinders and disk-shaped firebrands, the models reported in the work of Anthenien et al. (2006) were employed. In these models, the effects induced by the oscillatory motions of the particles were neglected, and the firebrands are assumed to always fall with their largest projected area normal to the relative wind velocity vector. Following the methodology applied for spherical firebrands, Anthenien et al.'s models for cylinders and disks were modified to allow their implementation in a three-dimensional turbulent flow field obtained from LES. The drag and gravitational forces are considered in the equation of linear momentum, leading to the general equations of particle motion previously presented for spherical particles in Eqs. (1) and (2).

For cylindrical firebrands, where a fixed aspect ratio $AR = 3$ was employed, the equation for the drag coefficient follows the work of Sucker and Brauer (1975):

$$C_d = 1.18 + \frac{6.8}{Re_d^{0.89}} + \frac{6.8}{Re_d^{0.89}} - \frac{4 \times 10^{-4} Re_D}{1 + 3.64 \times 10^{-7} Re_d^2} \tag{11}$$

This model is reported to be accurate over the range of $10^{-4} < Re_d < 2 \times 10^5$, a condition that is satisfied in all simulations performed here. The burning model for cylindrical particles is similar to the one employed for spherical particles, and the diameter regression rate assumes the form presented in Eq. (5):

$$\frac{d(d_p^2)}{dt} = -\beta \tag{12}$$

where $\beta$ ( $= \beta_o \left(1 + 0.276 Re_d^{1/2} Pr^{1/3}\right)$ ) is the burning parameter and is calculated in the same way for all firebrand shapes. Since the evaluation of the impact of the firebrand shape presented in Sect. 3.5 is performed for non-charring particles only, the burning models of spheres, cylinders, and disks simplify to the form presented in Eq. (12), where the diameter of the pyrolysis front equals the firebrand diameter ($d_{pyr} = d_p$). In the case of cylinders, the particle diameter is calculated as

$$d_p = \left(\frac{6 m_p}{\rho_p \pi}\right)^{1/3} \tag{13}$$

The calculation of the Nusselt number for cylindrical firebrands follows the relation presented by

Kramers (1946) for a cylinder in a gas flow perpendicular to its axis:

$$Nu_d = 0.42 Pr^{0.2} + 0.57 Pr^{1/3} Re_d^{1/2} \qquad (14).$$

For disk-shaped firebrands, the drag coefficient is calculated using the equation reported by Pitter et al. (1973) and Clift et al. (1978):

$$C_D = \left(\frac{64}{\pi Re_d}\right)(1 + 0.138 Re_d^{0.792}), \qquad Re_d \leq 130,$$

$$C_D = 1.17, \qquad Re_d > 130 \qquad (15).$$

Similar to the non-charring spheres and cylinders, the diameter regression rate of disk-shaped firebrands follows Eq. (12). The mass loss rate of the disk is expressed as

$$\frac{dm_p}{dt} = \frac{\pi \rho \tau}{4} \frac{d(d_p^2)}{dt} \qquad (16),$$

where $\tau$ ($= AR \times d_p$) is the thickness of the disk. For calculating the Nusselt number for disks, Eq. (14) is employed, which yields a good agreement with the results presented by Anthenien et al. (2006) for the disk-shaped firebrands.

A. Ganteaume, M. Guijarro, M. Jappiot, C. Hernando, C. Lampin-Maillet, P. Pérez-Gorostiaga, J. A. Vega (2011). **Laboratory characterization of firebrands involved in spot fires**. Annals of Forest Science, 68, 531-541.

Grimmond, T. R. Oke (1999). **Aerodynamic properties of urban areas derived from analysis of surface form**. Journal of Applied Meteorology and Climatology, 38, 1262-1292.

D. Guo, P. Zhao, R. Wang, R. Yao, J. Hu (2020). **Numerical simulations of the flow field and pollutant dispersion in an idealized urban area under different atmospheric stability conditions**. Process Safety and Environmental Protection, 136, 310-323.

J. E. Hilton, J. J. Sharples, N. Garg, M. Rudman, W. Swedosh, D. Commins (2019). **Wind-Terrain Effects on Firebrand Dynamics**. 23rd International Congress on Modelling and Simulation, Canberra, ACT, Australia 7.

K. Himoto, T. Tanaka (2005). **Transport of Disk-shaped Firebrands in a Turbulent Boundary Layer**. Fire Safety Science–Proceedings of The Eighth International Symposium, 433-444.

A. Inagaki, M. Kanda (2010). **Organized structure of active turbulence over an array of cubes within the logarithmic layer of atmospheric flow**. Boundary-Layer Meteorol, 135, 209-228.

A. Inagaki, M. C. L. Castillo, Y. Yamashita, M. Kanda, H. Takimoto (2012). **Large-eddy simulation of coherent flow structures within a cubical canopy**. Boundary-Layer Meteorol., 142(2), 207-222.

F. P. Incropera, D. P. DeWitt (1990). **Fundamentals of heat and mass transfer**. 3rd ed. New York, Wiley.

M. Kanda, R. Moriwaki, F. Kasamatsu (2004). **Large-eddy simulation of turbulent organized structures within and above explicitly resolved cube arrays**. Boundary-Layer Meteorol., 112, 343-368.

J. J. Kim, J. J. Baik (2010). **Effects of street-bottom and building-roof heating on flow in three-dimensional street canyons**. Advances in atmospheric sciences, 27(3), 513-527.

E. Koo, R. Linn, P. Pagni, C. Edminster (2012). **Modelling firebrand transport in wildfires using HIGRAD/FIRETEC**. International Journal of Wildland Fire, 21, 396-417.

H. Kramers (1946). **Heat transfer from spheres to flowing media**. Physica, 12, 61-80.

J. G. M. Kuerten (2016). **Point-Particle DNS and LES of Particle-Laden Turbulent flow - a state-of-the-art review**. Flow Turbulence Combust., 97, 689-713.

S. L. Lee, J. M. Hellman (1970). **Firebrand trajectory study using an empirical velocity-dependent burning law.** Combustion and Flame, 15(3), 265-274.

M. O. Letzel, M. Krane, S. Raasch, (2008). **High resolution urban large-eddy simulation studies from street canyon to neighbourhood scale**. Atmospheric Environment, 42(38), 8770-8784.

M. O. Letzel, C. Helmke, E. Ng, X. An, A. Lai, S. Raasch (2012). **LES case study on pedestrian level ventilation in two neighbourhoods in Hong Kong.** Meteorol. Z, 21(6), 575-589.

K. W. Lo, K. Ngan (2015). **Characterising the pollutant ventilation characteristics of street canyons using the tracer age and age spectrum**. Atmospheric Environment, 122, 611-621.

S. L. Manzello, T. G. Cleary, J. R. Shields, A. Maranghides, W. Mell, J. C. Yang (2008). **Experimental investigation of firebrands: Generation and ignition of fuel beds.** Fire Safety Journal, 43, 226-233.

S. L. Manzello, A. Maranghides, J. R. Shields, W. E. Mell, Y. Hayashi, D. Nii (2009). **Mass and size distribution of firebrands generated from burning Korean pine (Pinus koraiensis) trees**. Fire and Materials, 33, 21-31.
37